\documentclass[12pt,twoside]{article}   
\usepackage{epsfig,times}  
\usepackage{color}

\begin{document}
\thispagestyle{empty} 


 \renewcommand{\topfraction}{.99}      
 \renewcommand{\bottomfraction}{.99} 
 \renewcommand{\textfraction}{.0}


\newcommand{\nc}{\newcommand}

\nc{\qI}[1]{\section{{#1}}}
\nc{\qA}[1]{\subsection{{#1}}}
\nc{\qun}[1]{\subsubsection{{#1}}}
\nc{\qa}[1]{\paragraph{{#1}}}

\def\qbu{\hfill \par \hskip 6mm $ \bullet $ \hskip 2mm}
\def\qee#1{\hfill \par \hskip 6mm #1 \hskip 2 mm}

\nc{\qfoot}[1]{\footnote{{#1}}}
\def\qL{\hfill \break}
\def\qpar{\vskip 2mm plus 0.2mm minus 0.2mm}
\def\tvi{\vrule height 12pt depth 5pt width 0pt}
\def\qtvi{\vrule height 2pt depth 5pt width 0pt}
\def\qth{\vrule height 15pt depth 0pt width 0pt}
\def\qtb{\vrule height 0pt depth 5pt width 0pt}

\def\qparr{ \vskip 1.0mm plus 0.2mm minus 0.2mm \hangindent=10mm
\hangafter=1}

\def\qdec#1{\par {\leftskip=2cm {#1} \par}}
%
\def\qbfb#1{{\bf\color{\blue}{#1} }}

\def\qdpt{\partial_t}
\def\qdpx{\partial_x}
\def\qddpt{\partial^{2}_{t^2}}
\def\qddpx{\partial^{2}_{x^2}}
\def\qn#1{\eqno \hbox{(#1)}}
\def\qds{\displaystyle}
\def\qw{\widetilde}
\def\qmax{\mathop{\rm Max}}   
\def\qmin{\mathop{\rm Min}}   

\def\qs#1{{\bf \color{blue} \LARGE {#1}}\quad }

\def\qv{\vskip 0.1mm plus 0.05mm minus 0.05mm}
\def\qhu{\hskip 1mm}
\def\qhv{\hskip 3mm}
\def\qvv{\vskip 0.5mm plus 0.2mm minus 0.2mm}
\def\qhw{\hskip 1.5mm}
\def\qleg#1#2#3{\noindent {\bf \small #1\qhw}{\small #2\qhw}{\it \small #3}\qv }


\centerline{\bf \Large  Impact of personal income on mortality by age:}
\vskip 2mm
\centerline{\bf \Large  biological versus socio-economic effects}
\vskip 5mm

\centerline{Peter Richmond$ ^1 $, Wonguk Cho$ ^2 $, Beom Jun Kim$ ^3 $, 
 Bertrand M. Roehner$ ^4 $}

\vskip 10mm

\centerline{Version of 3 March 2021}
\vskip 10mm

{\small Key-words: income, age-specific death rate, infant, elderly}

\vskip 10mm

{\bf Abstract} \qL
The influence of  per capita income on life expectancy
is well documented, mostly through studies
of multinational samples. 
However, one expects fairly weak correlations
at both ends of the life span, that is to say in
early infancy and in age groups of elderly from 85 to 100 years.
The reason is that at both ends  mortality is largely
controled by biological factors rather than by socio-economic
conditions.\qL
In order to test this conjecture, we explore the influence 
of income on age groups, separately in France, 
the United States and South Korea.
More precisely in each country we compare income and mortality data 
in as many regional subunits as possible. 
One noteworthy constatation
is that, contrary to a common view, 
personal income is only weakly correlated with infant
mortality (i.e. mortality under the age of one year).
More broadly, we propose as a conjecture that
the common pattern revealed by 
the analysis of the three countries is
also valid in other developed countries.

\vskip 10mm

1: School of Physics, Trinity College Dublin, Ireland. \qL
Email: peter\_richmond@ymail.com
\qpar

2: Physics Department, Sungkyunkwan University, Seoul, South Korea.\qL
Email: wonguk.cho@gmail.com
\qpar

3: Physics Deparment, Sungkyunkwan University, Seoul, South Korea.\qL
Email: beomjun.kim@gmail.com
\qpar

4: Institute for Theoretical and High Energy Physics (LPTHE),
Pierre and Marie Curie Campus, Sorbonne University, National
Center for Scientific Research (CNRS), Paris, France. \qL
Email: roehner@lpthe.jussieu.fr

\vfill \eject

\qI{Introduction}

Let us start by indicating one of the main
motivations of the present study. 

\qA{The case of infant mortality}

Infant mortality is often used as an
indicator of poverty and difficult living conditions.
The idea behind that is probably that newborns under
one year are fragile and therefore more affected by
hardships than older children. We wanted to see
whether this belief is supported by solid evidence.
\qpar

The curves of the correlation of infant death rate
with income 
which will be given in the paper show that this
argument is not correct. Under one year of age
the correlation is rather low and not
statistically significant. This is understandable because
under one year
the causes of death are mostly due to congenital anomalies
which means that they 
are of biological rather than social origin.

\qA{Influence of  age  in the income-mortality relationship}

The studies published in recent times on the relation
between mortality and personal income 
(Marmot 2005, Cutler et al. 2006, Berkman et al. 2014,
Chetty et al. 2016) are not much concerned with how 
this relation changes with age which is the topic
of the present note.
\qL
The only studies of this question seem to go back to
the 1960s; for instance,
Frederiksen (1966a,b) studied this point 
as an aside of a broad study of 
population growth.
\qpar

Apart from the specific issue of infant mortality
just mentioned what led us to 
investigate this effect of age is our awareness
of the special character of mortality at the
two ends of the age spectrum. This point is developed
in the next subsection.

\qA{Biological versus social factors}

Mortality is influenced by many factors, whether
biological (e.g. genetic background), medical (e.g. vaccines,
antibiotics), social
(e.g. marital status, see Richmond et al. 2016);
economic and cultural aspects are of importance
through what is globally called 
the standard of living. 
\qpar

Among all these conditions medical factors are
of primary importance but
a previous investigation (Richmond et al. 2016a)
had led us to the observation that medical progress
has little influence on mortality at the two ends of
life span, i.e. on
very young and very old age groups, more precisely
during 
the first days or weeks after birth and similarly in
ages over 90. For these age groups
mortality is dominated by congenital
malformations on the young end and by a multitude
of wear-out effects which plague almost all organs
in old age. In short, the mortality of these
age groups appears largely determined by purely
biological factors. Therefore one does not expect
any substantial impact of other factors
and this includes economic factors.
Testing this conjecture was the main incentive for
embarking on the present study.
\qpar

If the conjecture is true,
since one knows that personal income has
a substantial influence on the life expectancy 
of adults (see for
instance Preston 1975, Fig.1) a natural
question was to explore the transitions between
adulthood and the weak impact regimes 
prior and after adulthood..

\qA{Answer to a possible objection}

In Preston (2007, p.484) one reads
that the correlation between infant mortality and
level of income is commonly found to be of the order
of -0.8. This shows a close relationship; actually it is
sufficiently strong for
infant mortality sometimes to be used as a proxy of income level.
How is this compatible with the low correlation mentioned
above? 
\qpar

The reason is very simple.\qL
Correlations as high as -0.8 are found in
cross-sectional studies involving a broad set of countries
ranging from developing to highly developed countries.
For such samples the correlation reflects 
the fact that in developed countries the
death toll of infectious diseases has almost vanished,
whereas in developing countries it still represents
a major cause of death in early infancy.
This difference is demonstrated very clearly by
the levesl of the annual infant mortality: itself, 
namely around 3 per 1,000
in advanced countries whereas  still 10 times 
more in many developing countries..

\qA{Multinational versus same nation samples}

In the present paper, we consider
more homogeneous data, namely
cross-regional data within three developed
countries, France, the United States and South Korea%
\qfoot{The fact that France is studied first is not because
one of the authors is French. The criterion
which was used is the number of regional subunits:
about 90 in France, 50 in the US and 15 in South Korea.
In principle more subunits should give more accurate results
but one must also make sure that there are enough 
individuals in each subunit so as to keep statistical
fluctuations under control.}%
.
\qpar

This methodology
leads to correlations which are lower than
with multinational samples
but they are also
more significant for our purpose. For instance, it
raises the question of the respective role of
pregnancy and conditions prevailing
after birth. 
Do streneous living conditions prevailing in poor families 
(e.g. hard work, noise, pollution) during pregnancy 
lead to more congenital malformations and in turn,
(at least for the most severe of them) to neonatal deaths?
In fact, it will appear that
in the first few weeks after birth there is only
a weak connection with income. This suggests that, at
least in developed countries, living conditions
during pregnancy do not markedly affect the development
of the embryo and fetus. In other words, {\it in utero}
biological development seems fairly well protected from
external factors. \qL
How is this 
result compatible with the well-known harm due to smoking,
drinking alcohol or taking drugs which affects
embryogenesis? Well, the low correlation with income suggests
that such hazardous behavior is only weakly income-dependent.

\qA{Outline of the paper}

The paper proceeds as follows.
\qbu Firstly, we consider the case of France.
For our purpose this country has some commendable
aspects. 
(i) Metropolitan France is divided into 96 administrative
units called ``departements''. Given the relatively
limited size of the country as compared
with countries like the United States or China
these units are fairly homogeneous areas.  
(ii) French demographic statistics give
detailed infancy mortality data starting
in the first days after birth and on the 
old age side there are data for all 5-year age-groups until
the age of 100. \qL
Thanks to such broad data sets it is possible to give
a fairly comprehensive view of the income effect.
(iii) For each of the three countries that we selected 
at least one of the co-authors is a native speaker.
It is true that
nowaday translation softwares have largely
removed the language barrier.
Nevertheless, the ability to read the
explanations and qualifications by ourselves made
things easier.
\qbu In the second part of the paper we
examine the case of the United States.
Our main goal
is of course to see if the typical pattern
observed for France will also be found in the US.
As subunits the most appropriate choice consisted in
the 50 states (plus the District of
Columbia). Counties would be too small especially
in the east and regions would be too inhomogenous.
In addition to the methodology used in part one
we propose what may be called
a fast methodology in which one concentrates
on ``extreme'' cases, that is to say on the states
which have the lowest and highest incomes. This
methodology is based on the realization that if it would
be possible to conduct this observation as an experiment
(in other words, as a prospective investigation)
it would not make sense to include in the sample
a large group
of some 30 states which have almost all the same income,
Such a group adds very little to the degree of
significance of the correlation.
\qbu South Korea is the third country that we examine.
What was our motivation for including South Korea 
into this study. After observing that our theoretical
argument was confirmed in the two cases of France and the 
US, we became convinced that similar confirmation
would be observed in similar countries
e.g. Germany or the UK. To consider an Asian
country was a more interesting challenge. Not only are
there clear cultural differences but also
a much more rapid pattern of development.
Will the same ``law'' hold despite these differences?
That was the question and the answer was ``yes''.

\qI{Age-specific income/mortality correlation in France}

Infancy refers usually to the period of a child's life before
it can talk.
\qpar
%
\begin{figure}[htb]
\centerline{\psfig{width=16cm,figure=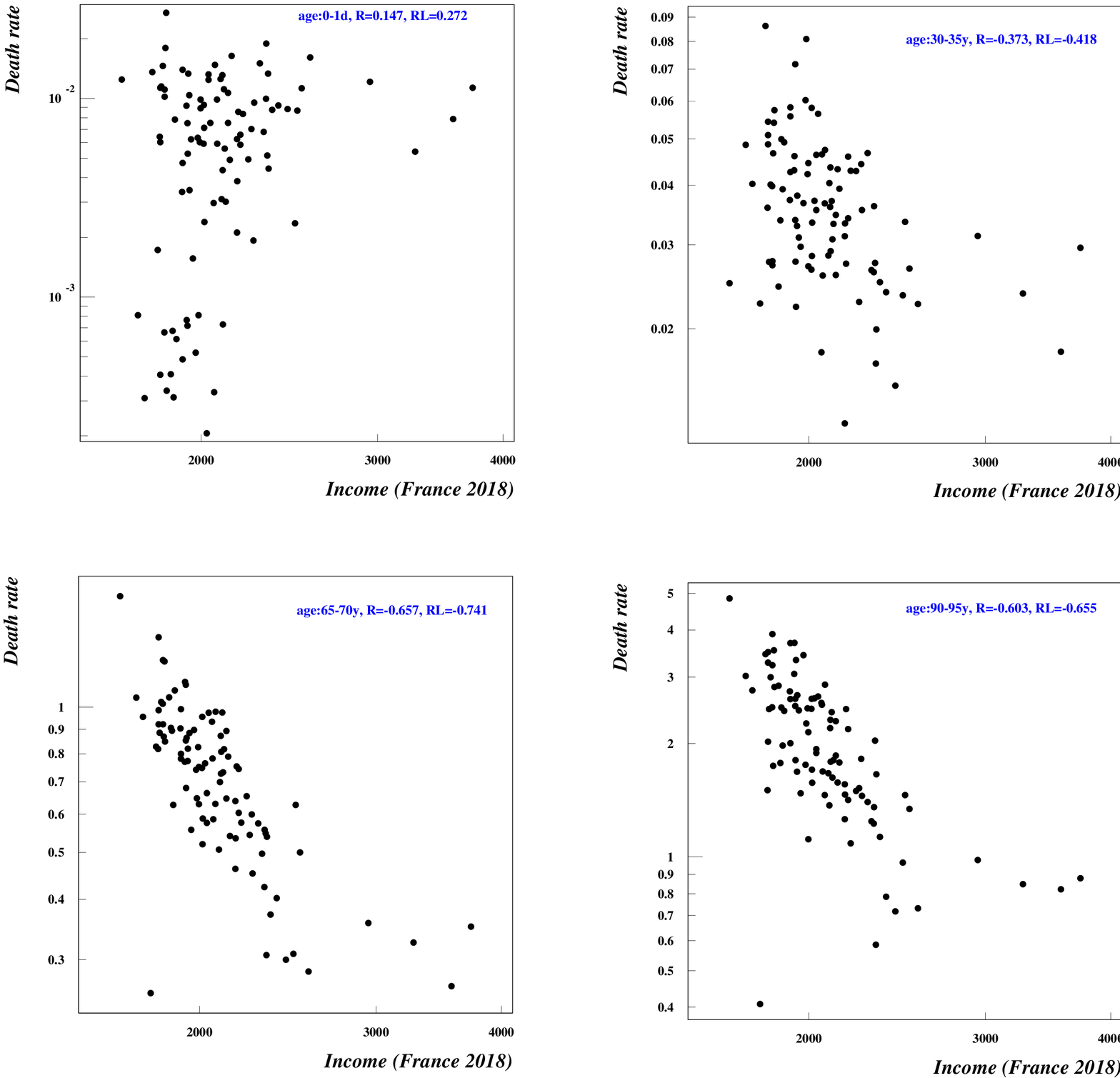}}
\qleg{Fig.1a,b,c,d \quad Relationship between income
and death rate in different age groups.}
{$ R $ gives the coefficient of linear correlation for
($ x= $income, $ y= $death rate) whereas $ RL $ gives the correlation
for the logarithms of the same variables.
Fig.1a shows the data points for the first age interval
whereas Fig.1d is for the last age interval.
Fig.1b,c which are for mid-age intervals show higher
correlations in conformity with life expectancy studies.}
{Sources: INSEE (Institut National de la Statistique
er des Etudes Economiques): DC1D, D\'ec\`es selon le sexe,
le groupe d'\^ages atteints dans l'ann\'ee.
Revenu mensuel d\'eclar\'e par foyer fiscal.}
\end{figure}

Here we give the word an extended meaning
so that it covers the age-interval during which the
age-specific death rate is decreasing  
This decrease is due to the progressive elimination
of new-borns affected by congenital anomalies.

Two comments are in order regarding Fig.1abcd.\qL
(i) Not surprisingly in most cases the mortality rate is a 
decreasing function of income. Whenever the correlation is positive
it is rather low and in fact not significant 9in the sense
that 0 belongs to the confidence interval of the correlation.
Because we are chiefly interested in the magnitude 
(i.e. the absolute value) of the correlation, it is the opposite
of the correlation which will be represented in subsequent graphs.\qL
(ii) THe relationship is not linear. When one draws the graphs
of Fig.1abcd in linear rather than in log-log axes one sees 
(not shown here) one sees clearly a steep fall for low and mid-income
which is followed by a part with a much flater slope.
On the log-log plot of Fig.1abcd the the relationship becomes 
almost linear with the second part being limited to a few data points.
%
\begin{figure}[htb]
\centerline{\psfig{width=8cm,figure=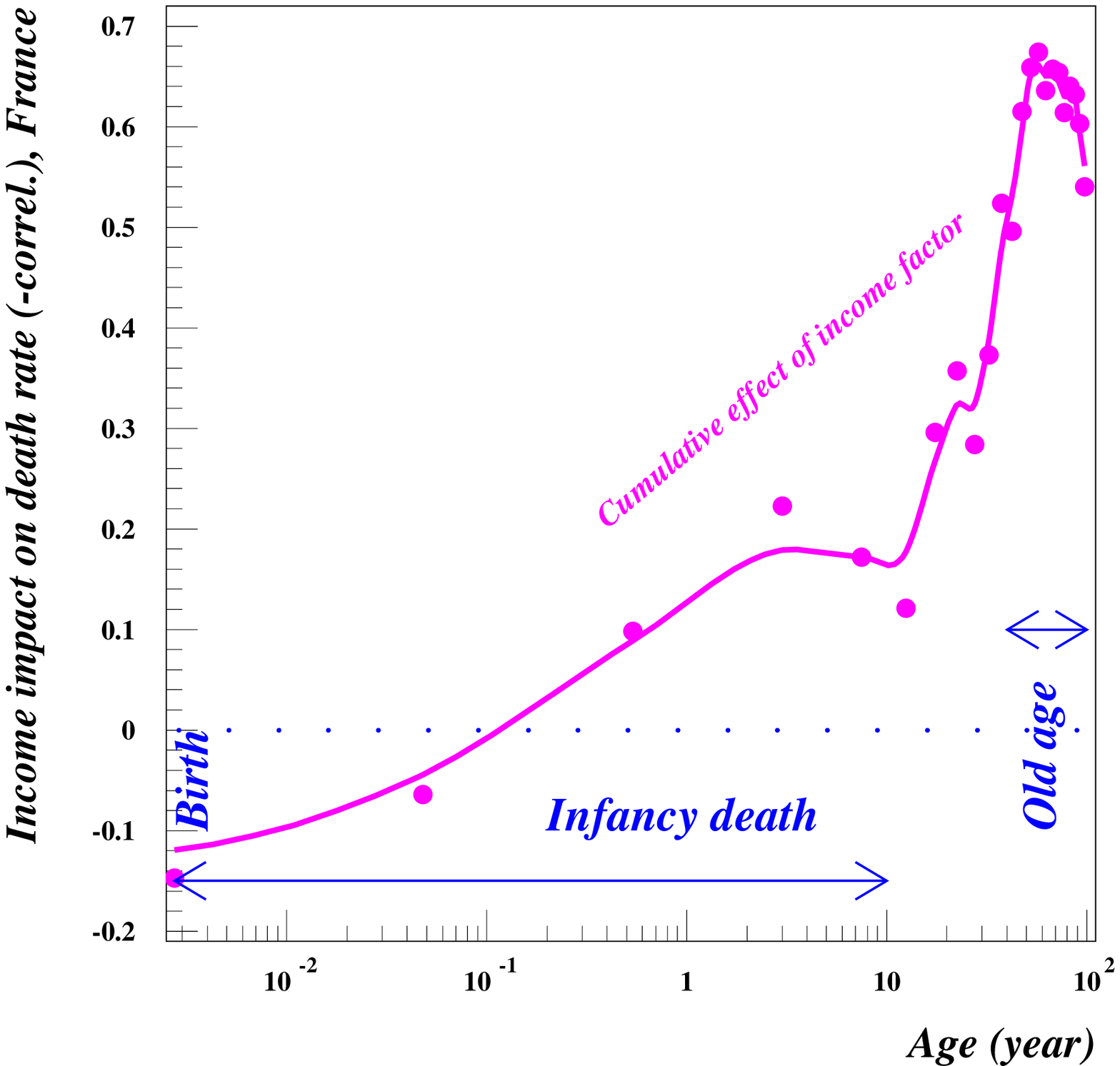}}
\qleg{Fig.2 \quad Income/mortality correlations as
a function of age in France, 2018}
{For convenience of the representation, the graph
shows the opposite of the correlations. 
The parts of the curve which are below the
zero line correspond basically to non significant
correlations.The red solid line is a centered moving
average over 3 successive data points, i.e.:
$ m(i)=[y(i-1)+2y(i)+y(i+1)]/4 $.}
{Sources: INSEE (Institut National de la Statistique
er des Etudes Economiques): DC1D, D\'ec\`es selon le sexe,
le groupe d'\^ages atteints dans l'ann\'ee.
Revenu mensuel d\'eclar\'e par foyer fiscal.}
\end{figure}

Correlations as a function of age are summarized in Fig.2.
\qpar

For elderly
the decrease of the correlation starts only
after the age of
90 and even then it is a fairly small fall.
In the United States we will see a more substantial fall.
Before one tries to find a definite reason one should observe 
that the mortality rates in age groups above 90 are beset with
large statistical fluctuations because of the 
small sizes of such age groups.
\qpar

Fig.2. shows the gradual build up of the
impact of economic factors. This part
of the graph answers the question
raised in the introduction about the transition
between the biological and socio-economic regimes. 
It is a gradual, not an abrupt, transition. 

\qA{Method}

ßefore we consider two other countries, let us summarize
the successive steps of the investigation.
\qee{1} The first, and possibly the most time-consuming step, is
to obtain the data. This is highly country-dependent. The
data may or may not be publicly available on Internet but
one can be sure that they have been recorded. This is fairly
evident for income because it is key-macroeconomic
variable. The same also holds, though for a different reason,
for age-specific mortality; the reason here is that there
are international agreements for the publication of death data
by causes of death. \qL
For the present investigation
such data are needed, not only nationaly, but for each of $ n $  
regional subunits, e.g. French d\'epartements or US states.
\qee{2} Then, for each of the age intervals, one plots the
$ n $ data points $ (x= $income, $ y= $death rate$ ) $ as shown in 
Fig.1a,b,c,d. At the same time one computes the correlation
between income and death rate.
\qee{3} Finally, the $ n $ correlations are plotted in a 
summary-graph whose $ x $ axis is age. This gives a graph
similar to Fig.2 on which one can read how the connection
between income and death changes with age.
.

\qI{The age-specific income/mortality relationship in the US}

In this section we examine the same variables
in the United States.

%
\begin{figure}[htb]
\centerline{\psfig{width=16cm,figure=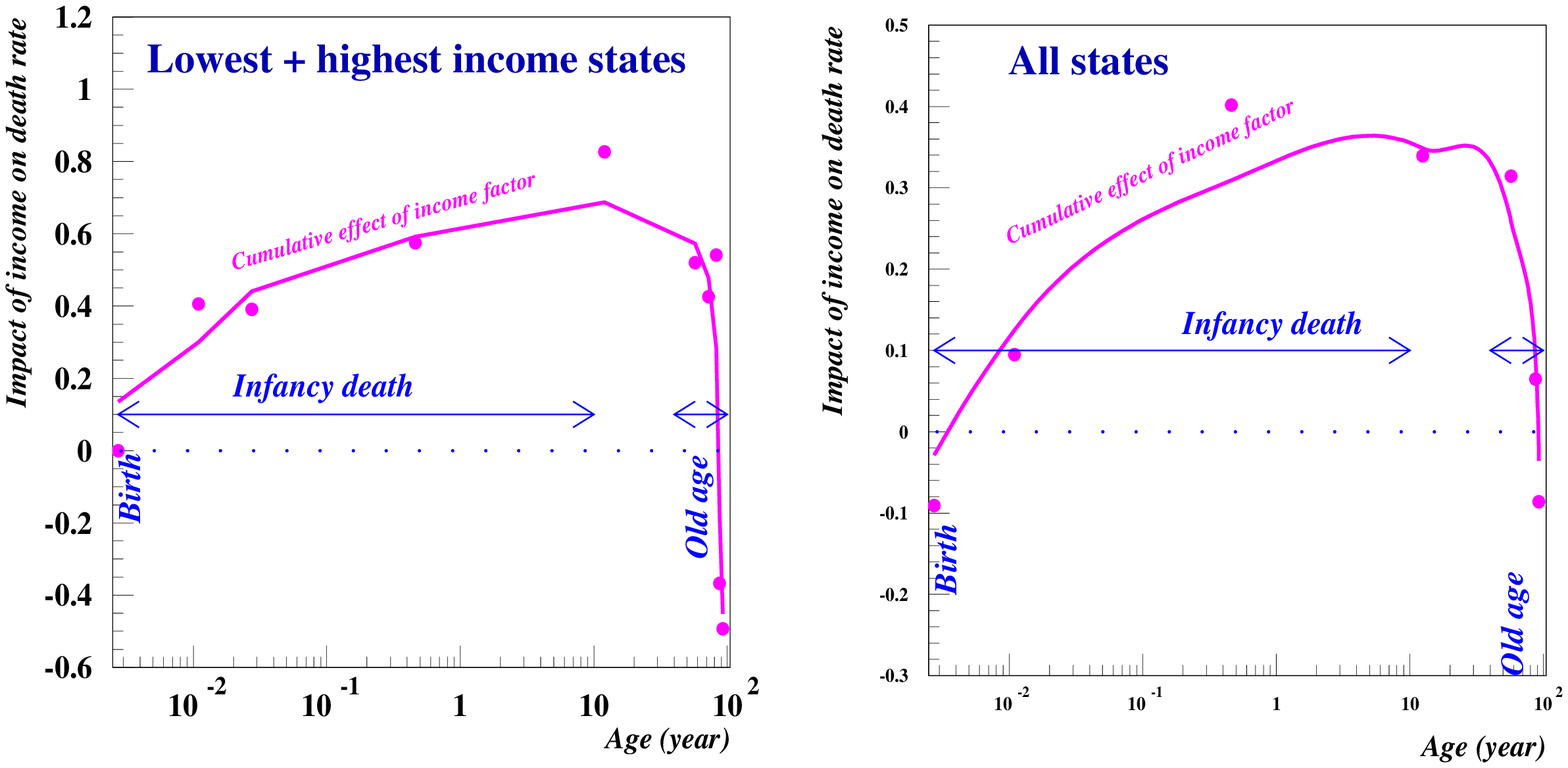}}
\qleg{Fig.3a,b \quad Income/mortality correlation as
a function of age, United States, 2010.}
{Left-hand side: graph in which the focus was
on the 5 states with lowest and highest incomes respectively;
right-hand side: graph for all states.}
{Sources: Personal income by state on the website
of the ``Bureau of Economic Analysis'' (BEA);
CDC (Centers of Disease Control) WONDER website
for detailed mortality by underlying causes
of death.}
\end{figure}

In the US the income data can be found on
the website of the ``Bureau of Economic Analysis''
whereas the mortality data are available on
the ``CDC-WONDER''  website for detailed 
underlying mortality.
\qpar

The analysis which leads to Fig.3a relies on the
repetition of the same observation in each state. 
Let us for a moment assume that one can 
choose at will the income level in each state. 
Then, the most accurate procedure would be to take 
as many incomes as possible in extreme
(i.e. very low
and very high) income levels and to cover the mid-income
interval with a number of cases providing uniform covering
density. 
This makes sense because the confidence interval of
a correlation is determined by two factors: (i) the
level of the correlation and (ii) the number of data points.
As one expects the correlations on both ends to be lower
than in the middle, a high density of data points will be
necessary to define accurately the left- and right-ends
of the curve. 
\qpar

How can one use this argument in the real situation 
where income levels cannot be attributed at will?
While it is of course impossible to
generate more data points than those provided,
one can 
limit the number of data points falling in the mid-income
range. The main advantage of this procedure is that
one can get good correlation estimates just by visual inspection
of the position of the lowest and highest
income cases. This procedure gives a clearer
insight into why correlations are low or high.
The comparison of Fig.3a and 3b shows that with only 10 states
one gets a picture very similar to the one based on 51 states.

\qI{The age-specific income/mortality relationship in South Korea}
   
    In case of South Korea, the data of regional income and mortality rate by
province can be obtained from the Korea National Statistical Office (KNSO).
However, there is no publicly open source for detailed infant 
mortality data by
province in Korea. Hence, we acquired the raw data of 
``The Complementary Survey
for Causes of Death'' from ``Microdata Integrated Service'' (MDIS) 
and processed the
data to calculate the infant mortality of different age periods by
province.
%
\begin{figure}[htb]
\centerline{\psfig{width=16cm,figure=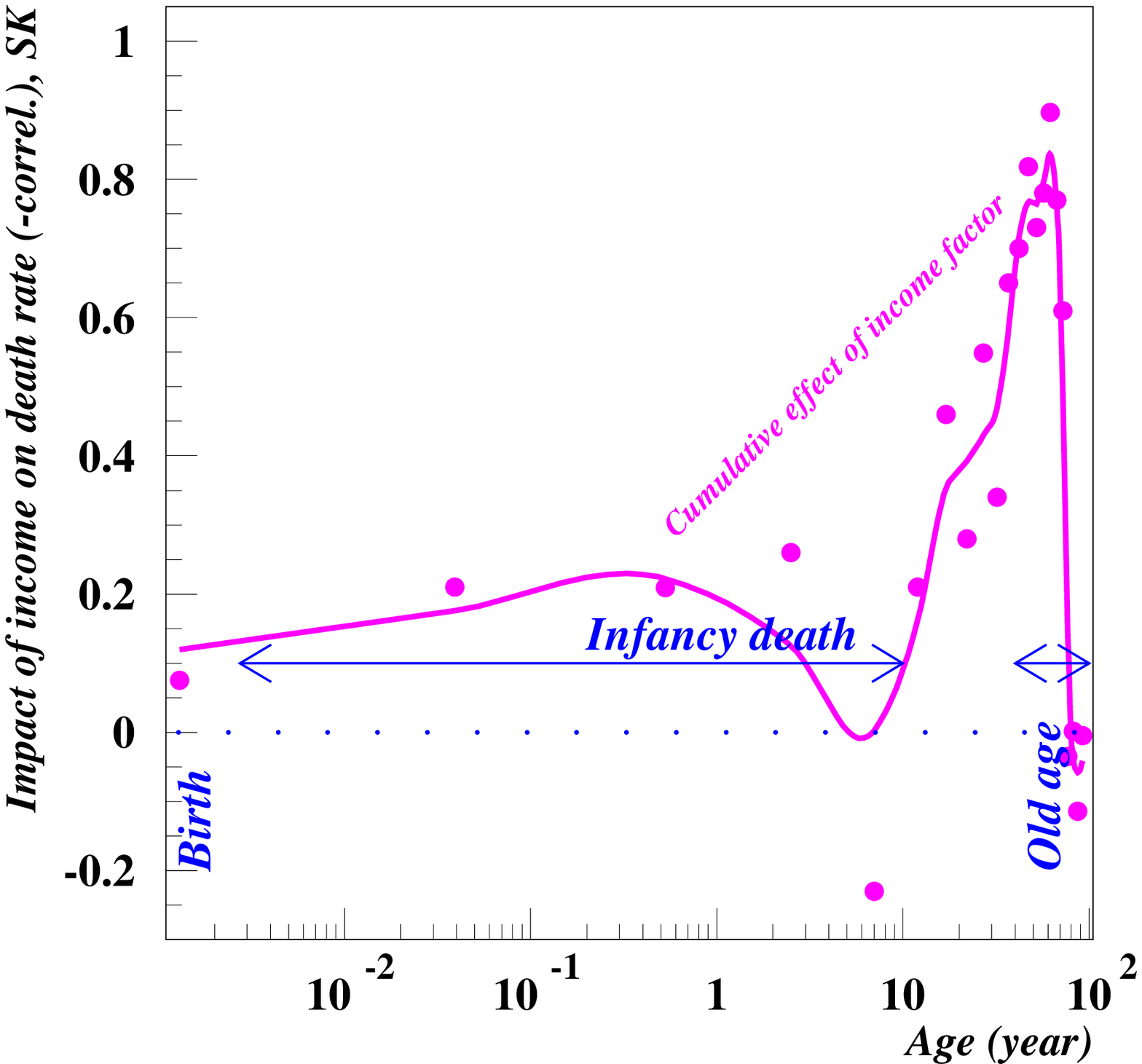}}
\qleg{Fig.4 \quad Income/mortality correlation as
a function of age, South Korea, 2017.}
{The data consist of 16 provincial level divisions of Korea.}
{Sources: Regional income and mortality data obtained from the 
Korea National Statistical Office (KNSO). 
Infant mortality data are produced based on the survey
data acquired from Microdata Integrated Service (MDIS).}
\end{figure}
    
 Fig. 4 displays the similar pattern with the previous cases. 
As expected, the
mortality rate in Korea shows little correlation with income at 
both ends of the
life span, where biological factors are more dominant 
than socio-economic ones.
Specifically, the effect of income factors gradually accumulates 
with age until
it reaches the peak at the age of 60-64 (r=-0.89), and a sharp fall 
occurs
between the groups of the age 70-74 ($ r=-0.61 $) and the age 
75-79 ($ r=-0.03 $).  This
observation confirms that our theoretical argument holds not only 
in the cases
of the Western countries but also in an Asian country despite 
their cultural difference.

\qI{Conclusion}

The methodology used in the present paper
differs from the more standard approach mainly in 
two ways.
\qbu Instead of using a sample comprising different
countries we analyzed regional units within the same
country.
In terms of health and mortality the main difference
between developing and developed countries is 
the fact that infectious diseases are still important causes
of death in the first whereas they have been
almost completely eliminated in the second.
For that reason one is not surprised to find
a high correlation between income and life expectancy
in samples
containing a mixture of developing and
developed countries.
The correlations analyzed here cover effects
which are more subtle.
\qbu Most publications mentioned in the introduction took
life expectancy as their target
variable but this does not allow to study separately
the effect of income level on diverse age groups%
\qfoot{Whereas it is possible to explore the
old-age interval by computing the life
expectancy for people having reached 60, 70 or 80 years, 
it is impossible to explore the infancy age
intervals in the same way.}
\qpar

What should be the next step? \qL
In sociology and in economics it is not so
common to be able to predict what will be observed. 
In a physics-like perspective the next step would be to check to
what extent a similar pattern holds in other developed
countries, for instance Australia, Canada, China, Germany, Italy,
Japan or  Spain. 
As in physics cases which do {\it not} follow the
law will be of greater interest than those which are
in agreement with it for then one must find out
for what reason the law is violated. This 
should bring about a better understanding.
\vskip 3mm

{\bf Ethical statement}
\qee{1} The authors did not receive any funding.
\qee{2} The authors do not have any conflict of interest.
\qee{3} The study does not involve any
experiment with animals that would require ethical approval.
\qee{4} The study does not involve any participants that
would have to give their informed consent.

\vskip 7mm 

{\bf References}

\qparr
Berkman (L.), Kawachi (I.), Glymour (M.) 2014: Social epidemiology. 
Oxford University Press, Oxford (England).

\qparr
Berrut (S.), Pouillard (V.), Richmond (P.), Roehner (B.M.) 2016:
Deciphering infant mortality.
Physica A 463,400-426.

\qparr
Chetty (R.), Stepner (M.), Abraham (S.), Lin (S.), Scuderi (B.),
Turner (N.), Bergeron (A.), Cutler (D.) 2016:
The association between income and life expectancy in the 
United States, 2001,2262014.
Journal of the American Medical Association (JAMA) 
315,16,1750,2261766.

\qparr
Cutler (D.), Deaton (A.), Lleras-Muney (A.) 2006:
The determinants of mortality. 
Journal of Economic Perspectives 20,97-120. 

\qparr
Fredericksen (H.) 1966: Dynamic equilibrium of
economic and demographic transition.
Economic Development and Cultural Change 14,316-322 (March 1966)

\qparr
Fredericksen (H.) 1966: Determinants and consequences of
mortality and fertility trends.
Public Health Reports 81,8,715-728 (August 1966)

\qparr
Marmot (M.) 2005: Social determinants of health inequalities. 
Lancet 365,9464,1099-1104. 

\qparr
Preston (S.H.) 1975: The changing relation between mortality
and level of economic development.
Population Studies 29,2,231-248.\qL
The paper was republished in 2007, under the following 
reference: 
International Journal of Epidemiology 36,484-490.

\qparr
Richmond (P.), Roehner (B.M.) 2016a: Predictive
implications of Gompertz's law. 
Physica A 447 446-454.

\qpar
Richmond (P.), Roehner (B.M.) 2016b: 
Effect of marital status on death rates. Part 1: High 
accuracy exploration of the Farr Bertillon effect.
Physica A 450,748-767. 

\end{document}